\documentclass[conference]{IEEEtran}
\IEEEoverridecommandlockouts
\usepackage{cite}
\usepackage{amsmath,amssymb,amsfonts}
\usepackage{algorithmic}
\usepackage{graphicx}
\usepackage{textcomp}
\usepackage{caption}
\usepackage[normalem]{ulem}
\usepackage{array}
\usepackage{changepage}
\usepackage[table,xcdraw]{xcolor}
\usepackage[normalem]{ulem}
\usepackage{array}
\usepackage{makecell}
\usepackage{graphicx}
\usepackage{rotating}
\usepackage{adjustbox}
\usepackage{colortbl}

\usepackage{tabularx}
\usepackage[normalem]{ulem}
\usepackage{array}
\def\BibTeX{{\rm B\kern-.05em{\sc i\kern-.025em b}\kern-.08em
    T\kern-.1667em\lower.7ex\hbox{E}\kern-.125emX}}
\begin{document}

\title{ Overcoming LLM Challenges using RAG-Driven Precision in Coffee Leaf Disease Remediation}

\author{\IEEEauthorblockN{Dr. Selva Kumar S}
\IEEEauthorblockA{\textit{Department of Computer Science and Engineering} \\
\textit{B. M. S. College of Engineering}\\
Bangalore, India \\
selva.cse@bmsce.ac.in} &

\IEEEauthorblockN{Imadh Ajaz Banday}
\IEEEauthorblockA{\textit{Department of Computer Science and Engineering} \\
\textit{B. M. S. College of Engineering}\\
Bangalore, India \\
imadh.cs20@bmsce.ac.in}  &

\IEEEauthorblockN{Vibha Venkatesh Shanbhag}
\IEEEauthorblockA{\textit{Department of Computer Science and Engineering} \\
\textit{B. M. S. College of Engineering}\\
Bangalore, India \\
vibha.cs20@bmsce.ac.in} \and

\IEEEauthorblockN{Afifah Khan Mohammed Ajmal Khan}
\IEEEauthorblockA{\textit{Department of Computer Science and Engineering} \\
\textit{B. M. S. College of Engineering}\\
Bangalore, India \\
afifah.cs20@bmsce.ac.in}&

\IEEEauthorblockN{Manikantha Gada}
\IEEEauthorblockA{\textit{Department of Computer Science and Engineering} \\
\textit{B. M. S. College of Engineering}\\
Bangalore, India \\
manikantha.cs20@bmsce.ac.in}

}

\maketitle

\begin{abstract}

This research introduces an innovative AI-driven precision agriculture system, leveraging YOLOv8 for disease identification and Retrieval Augmented Generation (RAG) for context-aware diagnosis. Focused on addressing the challenges of diseases affecting the coffee production sector in Karnataka, The system integrates sophisticated object detection techniques with language models to address the inherent constraints associated with Large Language Models (LLMs). Our methodology not only tackles the issue of hallucinations in LLMs, but also introduces dynamic disease identification and remediation strategies. Real-time monitoring, collaborative dataset expansion, and organizational involvement ensure the system's adaptability in diverse agricultural settings. The effect of the suggested system extends beyond automation, aiming to secure food supplies, protect livelihoods, and promote eco-friendly farming practices. By facilitating precise disease identification, the system contributes to sustainable and environmentally conscious agriculture, reducing reliance on pesticides. Looking to the future, the project envisions continuous development in RAG-integrated object detection systems, emphasizing scalability, reliability, and usability. This research strives to be a beacon for positive change in agriculture, aligning with global efforts toward sustainable and technologically enhanced food production.

\end{abstract}

\begin{IEEEkeywords}
YOLOv8, LLM, GPT-3.5, RAG, Precision Agriculture, Object Detection, NLP
\end{IEEEkeywords}

\section{Introduction}
In precision agriculture, the incorporation of cutting-edge technologies is essential for tackling challenges in disease identification and remediation. One such advancement is YOLO (You Only Look Once) which is an object detection algorithm that analyzes entire images at once, predicting both bounding boxes and class probabilities in a single pass. It employs anchor boxes to efficiently detect objects of different sizes and shapes. By utilizing a grid-based approach and a deep convolutional neural network backbone, YOLO achieves real-time performance and high accuracy across diverse domains. Through multiple iterations, YOLO has undergone improvements in speed, accuracy, and reliability, catering to applications such as autonomous driving and surveillance. The original YOLOv1, YOLOv2 (YOLO9000), YOLOv3, YOLOv4, and YOLOv5 were the first in the line of YOLO (You Only Look Once) object detection models. The most recent models in the line are YOLOv6, YOLOv7, and YOLOv8. Every iteration has sought to improve speed, precision, and adaptability for a range of uses, such as plant disease detection. Because of their real-time processing capabilities, these YOLO models have shown to be superior to conventional machine learning and deep learning techniques, making them indispensable tools for early intervention and disease mitigation in agriculture. YOLOv8 is an efficient object detection system renowned for its real-time identification of plant diseases. YOLOv8's speed, processing images in a single pass, proves invaluable for timely disease detection, particularly important in the dynamic landscape of agriculture. 
Despite YOLOv8's effectiveness in object detection, limitations arise with Large Language Models (LLMs) like GPT-3.5. These models, while powerful, exhibit a static nature, leading to inaccuracies in disease diagnosis, known as 'hallucination.' In agriculture's ever-changing conditions, the static nature of LLMs impedes their capability to provide accurate and up-to-date information, especially in contexts like the coffee industry, where various factors influence disease outcomes.

Through our paper, we propose a solution that addresses the issues mentioned by conducting a comprehensive review of Large Language Models (LLMs) and their drawbacks. Our aim is to reduce the research gap through a novel approach that mitigates the limitations of LLMs, particularly in the context of precision agriculture. To address these limitations, our research introduces Retrieval Augmented Generation (RAG), which, when integrated with LLMs like GPT-3.5, mitigates drawbacks by fetching up-to-date, context-specific data from external databases. Serving as a dynamic bridge, RAG minimizes the risk of hallucination, enhancing GenAI application accuracy by incorporating current, domain-specific knowledge. This approach ensures our model remains informed and significantly improves adaptability and reliability in providing context-aware solutions for precision agriculture. RAG's role in overcoming LLM limitations is central to our research, promising a transformative impact on agricultural practices.

\begin{figure*}[htbp]
\centerline{\includegraphics[width=10cm]{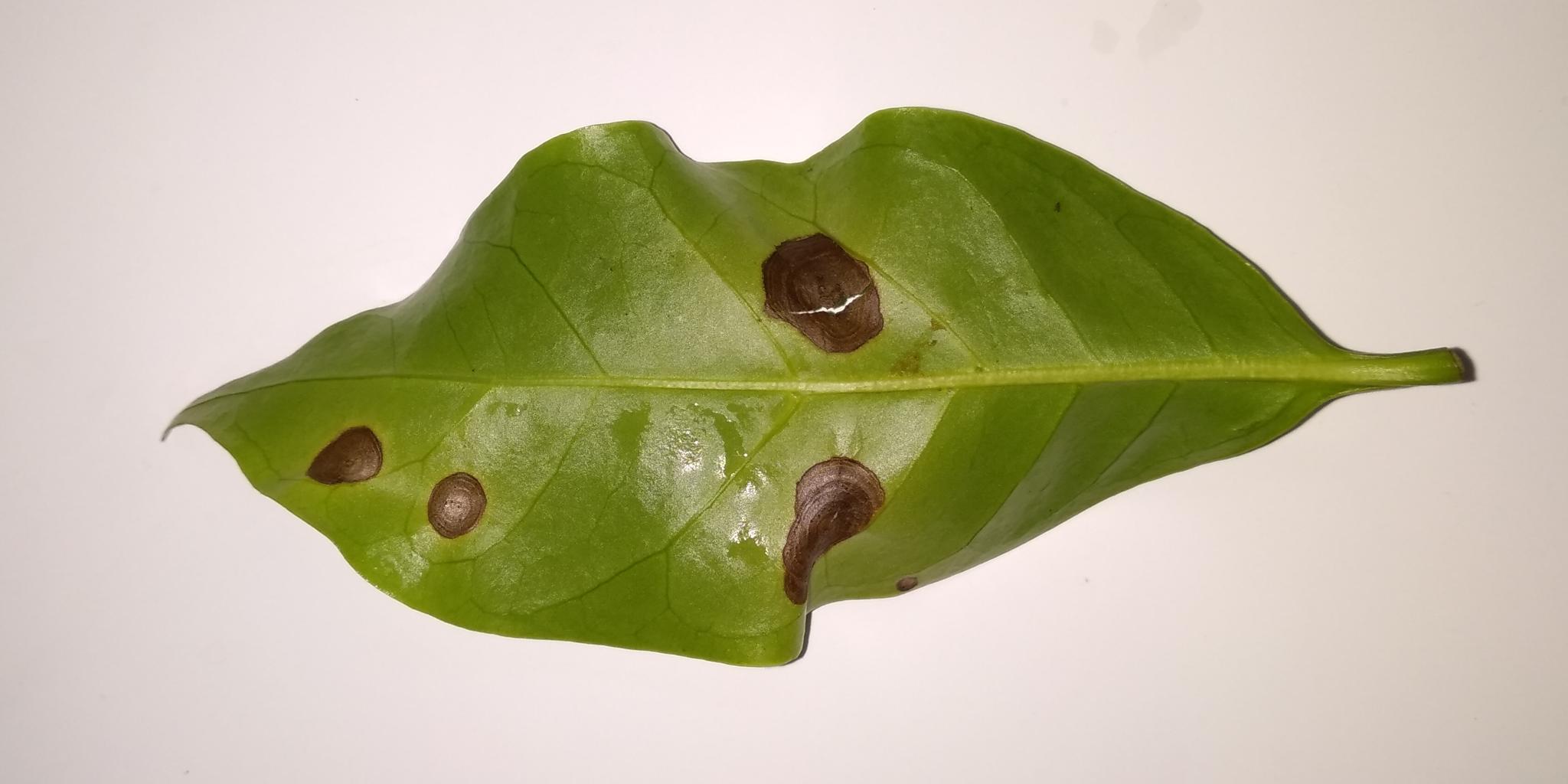}}
\caption{Diseased Leaf (Phoma)}
\label{fig1}
\end{figure*}

\section{Literature Survey}

The study\cite{b1} utilizes YOLOv3 for identifying plant diseases, leveraging its effective object detection capabilities. By training the model on labeled images of both diseased and healthy plant parts, the model attains high accuracy in detecting prevalent Apple tree diseases such as apple scab, cedar apple rust, and black rot. The next research\cite{b2} employed YOLOv5 for detecting rice leaf diseases, achieving notable precision (0.83), recall (0.94), and mAP (0.62) scores. YOLOv5 is distinguished by its rapid object detection model with enhanced performance, incorporating a backbone (CSPDarknet), neck (PANet), and head (Yolo Layer) architecture for effective detection. Wanx, X. et al.\cite{b3} proposed the the YOLO-Dense network, inspired by DenseNet that employs dense connections to boost feature extraction and enhance detection accuracy in identifying anomalies in tomatoes. This innovative method achieved an impressive mean average precision (mAP) of 96.41\%, demonstrating superior performance compared to other algorithms such as SSD, Faster R-CNN, and the original YOLOv3. Sajitha P et al., \cite{b4} introduced a system that integrates YOLO v7 for leaf disease detection and GPT-3 for providing corrective measures. And Jiajun Qing et al. in their paper \cite{b5},  combine the logical reasoning capabilities of GPT-4 with YOLOPC, a lightweight YOLO variant, to achieve real-time agricultural disease diagnosis. While YOLOPC attains 94.5\% accuracy with 75\% fewer parameters, GPT-4 demonstrates 90\% reasoning accuracy in generating diagnostic reports. Both \cite{b4} and \cite{b5} face the limitation that the language model might not consistently yield the most accurate answers. 

Jean Kaddour et al., \cite{b6} explore the challenges with LLMs through their paper “Challenges and Applications of Large Language Models”. The paper explores large language models (LLMs), highlighting challenges like intricate datasets and elevated expenses. It focuses on enhancing LLM behavior and knowledge, discusses fine-tuning methods, and emphasizes the need for comprehensive evaluation. In \cite{b7}, the authors explore the matter of hallucination in LLMs and propose the LVLM Hallucination Revisor (LURE), an algorithm designed to address object hallucination in Large Vision-Language Models (LVLMs). Evaluation on six open-source LVLMs demonstrates a substantial 23\% improvement in general object hallucination metrics compared to prior approaches. Further Junyi Li et al., through their paper \cite{b8} introduce HaluEval which is a comprehensive collection of 35,000 hallucinated and normal samples for analyzing and evaluating LLMs. The study presents a two-stage framework for generating and annotating hallucinated samples, revealing that existing LLMs often fail to recognize hallucinations in text and tend to generate such content. \cite{b9} provides an overview of the challenges posed by hallucination in LLMs. It delves into the impact of noisy data during pre-training on LLMs' parametric knowledge and the subsequent occurrence of hallucinations. The survey explores various mitigation strategies, including data curation, filtering, and supervised fine-tuning, as well as the use of high-quality reference corpora. In summary, \cite{b7}, \cite{b8}, and \cite{b9} collectively reveal that hallucination is a pervasive challenge in LLMs, prompting the exploration of algorithms, benchmarks, and mitigation strategies to enhance their performance and reliability.

In \cite{b10} the authors address hallucination in Language Models through the creation of the HILT dataset, utilizing 75,000 text passages generated by 15 LLMs. They emphasize the need for continuous updates due to the evolving nature of the field, and present detailed statistics on factual mirage (FM) and silver lining (SL) categories. Nitin 
Liladhar Rane et al. \cite{b11} explore the multifaceted contributions of large language models, such as ChatGPT, in scientific and research progress across diverse domains. It underlines the potential for these models to revolutionize knowledge dissemination while acknowledging the ethical and societal implications and emphasizing the importance of responsible development, deployment, and regulation.
In \cite{b12} Abdullahi Saka et al. shift the focus to the utilization of GPT models in the construction industry. The authors delve into opportunities, limitations, and a use case validation involving NLP techniques for processing construction project documents and Building Information Modeling (BIM) data. Abdullahi Saka et al. suggest formulating ethical use policies, exploring novel applications, and researching solutions for GPT model limitations in construction.

\begin{figure*}[htbp]
\centerline{\includegraphics[width=10cm]{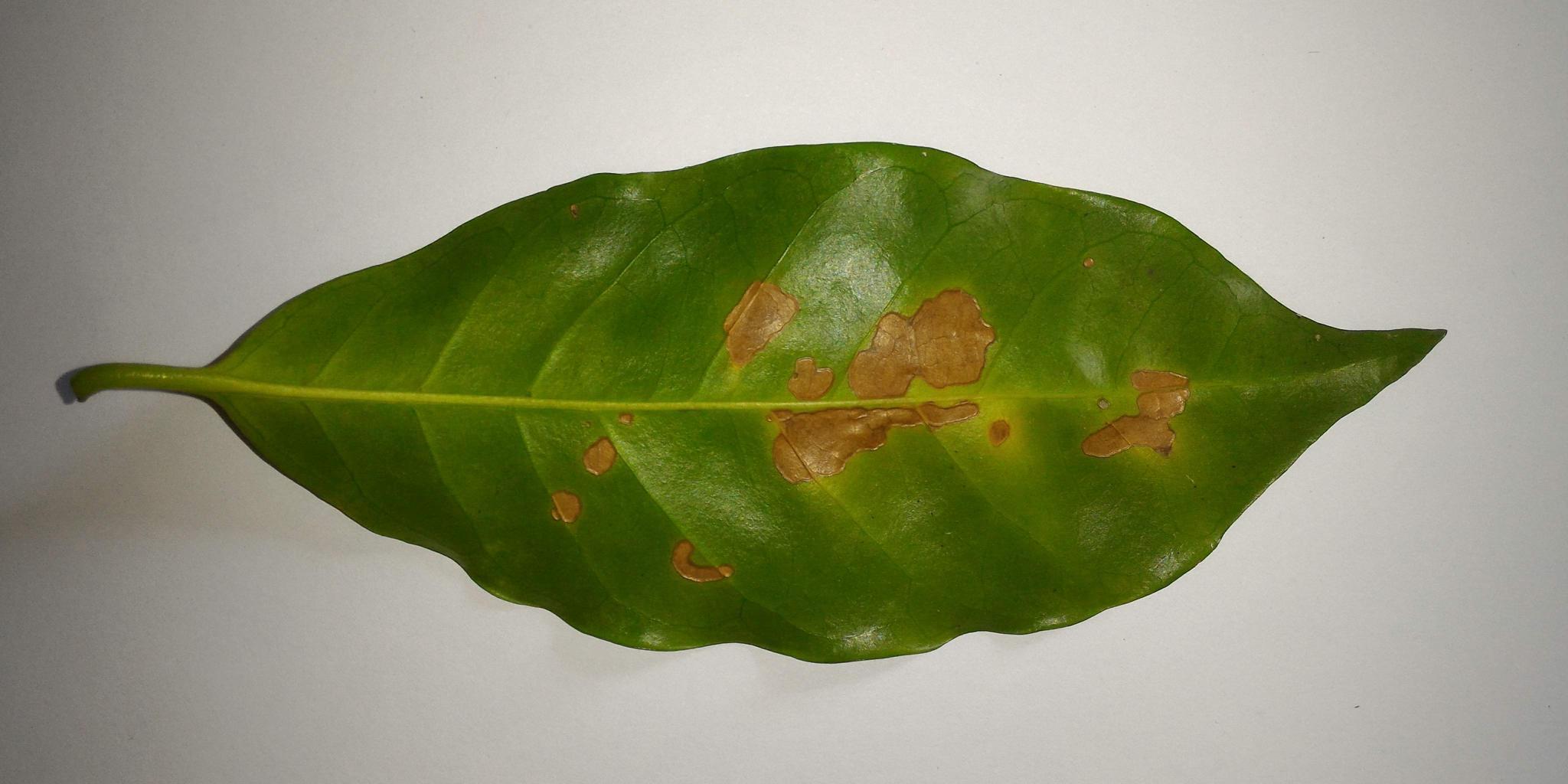}}
\caption{Diseased Leaf (Miner)}
\label{fig1}
\end{figure*} 

The paper \cite{b13} presents a balanced assessment of large language models (LLMs). It highlights the sophisticated inductive learning and inference capabilities of LLMs, including their capacity to recognize hierarchical syntactic structure and complex semantic relations. Additionally, LLMs have demonstrated potential in tasks such as medical image analysis and diagnostics and predicting the properties of proteins and new molecular structures. However, Shalom Lappin also acknowledges limitations, such as the potential for LLMs to hallucinate plausible-sounding narratives with no factual basis, their susceptibility to adversarial testing, and the requirement for additional investigation into improving their performance on specific tasks, addressing biases, and developing smaller, more lightweight models.

\cite{b14} Introduces the Retrieval-Augmented Generation (RAG) model, showcasing its state-of-the-art performance in open-domain question answering tasks, emphasizing its ability to generate diverse and factual content. The next paper \cite{b15}, Self-RAG, presents a framework that enhances large language models (LLMs) through retrieval and self-reflection without compromising creativity. It employs instruction and demonstration pairs, achieving significant improvements in model performance, factuality, and citation accuracy, with future work aiming to address factual errors and enhance self-reflection mechanisms. The authors of \cite{b16} examine the incorporation of extensive language models into systems for retrieving information, highlighting potential benefits and challenges, proposing research directions for improvement, and addressing drawbacks such as bias and data requirements. Jiawei Chen et al. \cite{b17} establish a standard for assessing the effectiveness of LLMs in retrieval-augmented generation tasks, identifying limitations in noise robustness and information integration abilities, and suggesting directions for improvement. \cite{b18} proposes ARM-RAG, a system leveraging RAG to enhance the problem-solving capabilities of LLMs, demonstrating superior performance by utilizing Neural Information Retrieval for reasoning chains derived from solving math problems and suggesting avenues for further enhancements. 

In \cite{b19} "InPars," introduces a method using large language models (LLMs) for few-shot labeled data generation in information retrieval (IR) tasks, demonstrating superior performance and emphasizing potential with domain-specific training data. Limitations include the lack of pretraining and limited suitability for non-neural retrieval algorithms. The next paper \cite{b20}, "Retrieval-based Evaluation for LLMs," proposes Eval-RAG, an approach to evaluating LLM-generated texts in the legal domain, outperforming existing methods in correlation with human evaluation and factual error identification. Future work includes refining Eval-RAG, exploring its applicability to other domains, and addressing potential limitations. The authors in \cite{b21}, "Retrieval Meets Long Context LLMs," compare retrieval-augmented language models (RAG) and long context LLMs, demonstrating RAG's significant performance improvement in Q\&A and summarization tasks. 
Future work aims to explore combined retrieval and long context LLMs for enhanced accuracy, with limitations not explicitly mentioned.

Zhangyin Feng et al. \cite{b22} introduces Retrieval-Generation Synergy Augmented Large Language Models, showcasing an iterative framework that significantly enhances the cognitive reasoning capacity of expansive language models (LLMs) for knowledge-intensive tasks, particularly answering questions in a broad range of domains. Through experiments on four datasets, the proposed method outperforms previous baselines, demonstrating improved LLM reasoning. \cite{b23} focuses on Interpretable Long-Form Legal Question Answering with Retrieval-Augmented Large Language Models. It presents an end-to-end methodology that leverages a "retrieve-then-read" pipeline, employing a retrieval-augmented generator (RAG) approach with LLMs. The authors fine-tune on a task-specific dataset and introduce the Long-form Legal Question Answering (LLeQA) dataset. The authors  highlight the positive aspects of this approach, emphasizing its potential for generating syntactically correct answers relevant to legal questions. The last paper \cite{b24}, Establishes Performance Baselines in Fine-Tuning, Retrieval-Augmented Generation, and Soft-Prompting for Non-Specialist LLM Users, explores ways to improve LLM performance for non-specialist users. It compares unmodified GPT 3.5, fine-tuned GPT 3.5, and RAG using a limited dataset and technical skill. RAG stands out as an effective strategy, outperforming fine-tuning and showcasing positive results within the framework of the LayerZero cryptocurrency bridging project. The paper discusses the accessibility of these techniques to non-technical users and emphasizes the positive impact of RAG on LLM performance.

\begin{figure*}[htbp]
\centerline{\includegraphics[width=21cm]{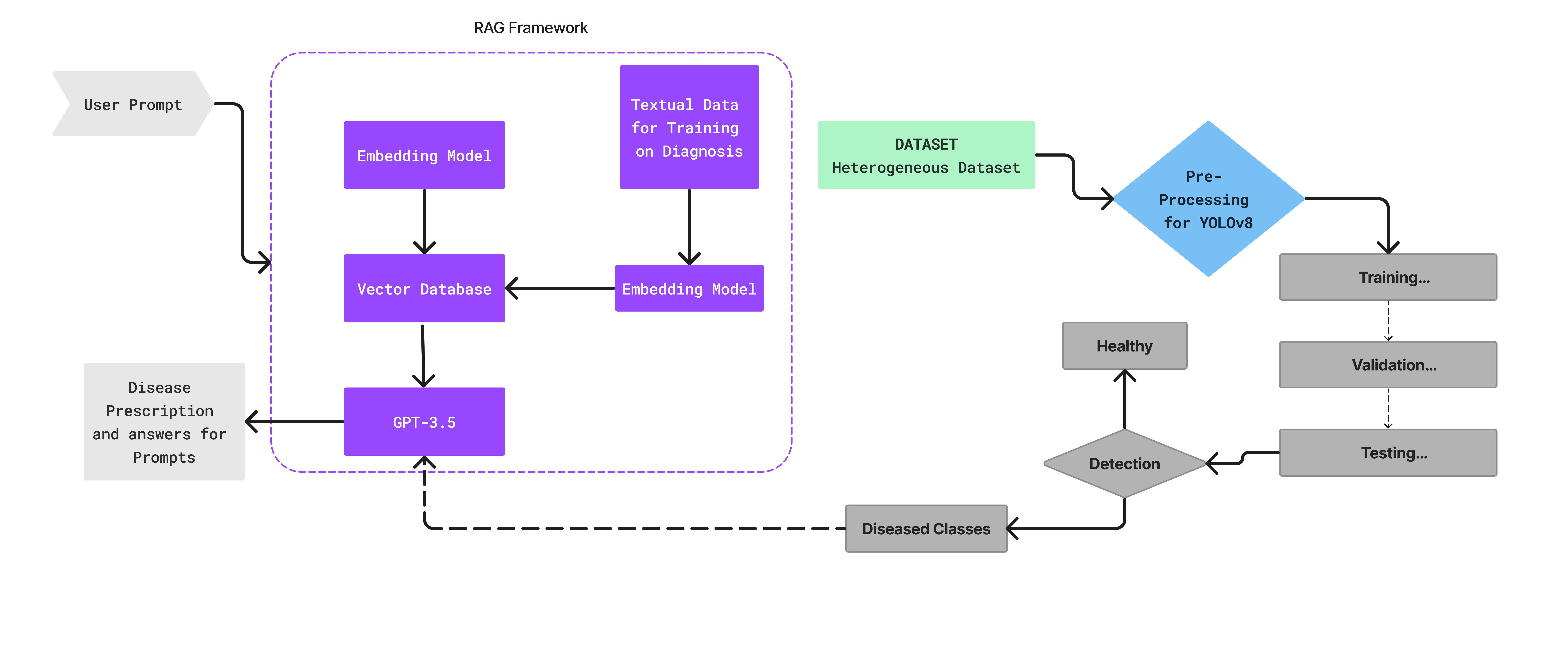}}
\caption{Proposed Methodology Workflow}
\label{fig1}
\end{figure*}

\captionsetup{skip=10pt}
\renewcommand{\arraystretch}{1.5}

\section{Methodology}

\subsection{Methodology Workflow}\label{AA}
In our methodology, illustrated in Fig. 4, for coffee leaf disease care and detection, we embrace a hybrid model. We train a YOLOv8 (You Only Look Once) Model for real-time instance segmentation on Coffee Leaves with the Transfer Learning approach, enhancing our ability to detect potential disease classes in coffee leaves such as Phoma and Miner as shown in figure 1 and 2 \cite{b25}. Simultaneously, we leverage the Natural Language understanding capabilities of GPT-3.5 (Generative Pre-Trained Transformer 3.5) to generate insightful diagnoses and treatment recommendations for the identified diseases. YOLOv8, developed by Ultralytics and pre-trained on Microsoft's COCO dataset, is a PyTorch-based model renowned for its modern techniques. It incorporates built-in Instance Segmentation, contributing to improved overall performance.

Additionally, we introduce the utilization of the Retrieval-Augmented Generation (RAG) framework, enhancing the role of GPT-3.5 in the generation process. RAG allows us to fetch up-to-date and context-specific data from external databases, providing valuable information to the language model during the generation of responses. This integration of RAG by building a robust pipeline from image capture to generating accurate results based on disease prescriptions aims to overcome limitations associated with static Large Language Models (LLMs) and ensure that our coffee leaf disease detection and care system remains adaptable, informed, and accurate in its recommendations.

\begin{itemize}
\item The workflow commences with the Phytopathology Recognition subsystem, where YOLOv8 processes input images, conducts plant disease identification, and classifies the findings. This involves utilizing the pre-trained YOLOv8 model for real-time instance segmentation and disease detection. The outcomes are then transmitted to the Remediation Assistance subsystem, where the Language Model (LLM), armed with the Retrieval-Augmented Generation (RAG) framework, generates highly contextual suggestions for effective remediation. The vector database becomes integral at this stage, storing and retrieving vector representations of remediation data to enhance the contextual accuracy of the suggestions.

\end{itemize}

\begin{itemize}
\item The Integration and Communication subsystem serves as a crucial bridge, establishing robust communication channels between the Phytopathology Recognition and Remediation Assistance components. It defines communication protocols, ensuring secure and reliable data exchange between these subsystems. The Decision Fusion Module operates as a nexus, receiving disease identification results from the Phytopathology Recognition subsystem and seamlessly integrating them with the remediation suggestions. This integration culminates in the generation of a comprehensive recommendation tailored for effective plant disease management. 
\end{itemize}
\begin{itemize}
\item The User Interface acts as the final frontier, providing users with a platform to interact with the system. Users can input plant images, receive detailed reports on identified diseases, and access suggested remediation measures. This interface facilitates user engagement and feedback, contributing to the continuous improvement of the system. The technical components and considerations encompass a range of elements, from communication interfaces to security measures, thereby shaping the overall advantages and disadvantages of the proposed system. 
\end{itemize}

\section{Experimental Setup}
\subsection{Dataset Compilation}
 We leverage a diverse dataset encompassing multiple sources from Kaggle \cite{b25}, open-source datasets, and real-time pictures of diseased coffee leaves obtained directly from coffee fields. 
 
\subsection{Dataset Annotation and Augmentation}
The YOLOv8 model relies on annotated datasets for effective training. Our dataset, sourced from various internet repositories and real-time images from coffee fields, undergoes thorough annotation with class labels (Rust, Miner, and Phoma). Annotated data is pre-processed, incorporating augmentation techniques like flipping, rotation, and brightness adjustment. This not only enhances dataset diversity but also mitigates the requirement for extra manual annotation efforts, streamlining the overall training process.

\subsection{YoloV8 Model}
The workflow starts by setting up the environment, including the installation of the required YOLOv8 library and dataset preparation. This involves checking and extracting the dataset, renaming necessary files, and organizing paths for efficient data management. Following setup, the training process is initiated, configuring model parameters and training settings. An image is uploaded for segmentation, and the model performs segmentation, saving the output. Lastly, a custom function is employed to display the segmented result, highlighting the versatile application of YOLOv8 across various segmentation tasks.

\subsection{Label Extraction}
The output from the YOLOv8 trained model is utilized to extract labels using the EasyOCR library. This process involves extracting text from the segmented image, filtering it to include relevant disease labels such as 'Rust', 'Miner', and 'Phoma', and subsequently printing the filtered output for analysis.

\subsection{Prompt Generation}
Unlike traditional object detection models, RAG-LLM requires a different form of input. We generate prompts from the dataset, encapsulating concise and relevant information about the coffee leaf diseases. These prompts act as queries to the language model for generating context-aware and detailed responses.

\subsection{Working of RAG-LLM}

After label extraction, the extracted labels serve as input, cross-referenced with the vector store. The disease diagnosis information contained in research PDFs act as the knowledge base. They undergo text extraction and segmentation which are converted into manageable chunks. These chunks contribute to the creation of a vector store generated by the OpenAI API embedding, optimizing storage for text embedding. A conversational chain is then established, allowing users to engage in dialogue and gather insights. This cohesive approach seamlessly integrates document information retrieval with AI-driven conversational capabilities, offering an efficient platform for exploring coffee diseases and their management strategies. Hence, allowing the information retrieval method proves out to be more efficient to this use case. The proposed RAG model is built using the Langchain framework to enable seamless integration with LLM powered Q\&A. 


\subsection{Challenges with the Dataset and Model Training}

In transitioning to RAG-LLM, challenges may arise in formulating effective prompts, ensuring the model's contextual understanding, and maintaining coherence in responses. Expert input remains crucial in refining the prompts and validating the model's outputs.

\section*{Conclusion}

This research represents a significant advancement in precision agriculture by integrating YOLOv8 and Retrieval Augmented Generation (RAG), offering a novel approach to disease identification and management in agriculture, with a focus on the coffee industry in Karnataka. While YOLOv8 excels in real-time disease detection, challenges persist with Large Language Models (LLMs) like GPT-3.5, particularly in addressing inaccuracies known as 'hallucination' due to their static nature. To overcome these limitations, our study introduces RAG, which dynamically retrieves context-specific data from external sources, enhancing the accuracy and adaptability of the GenAI application. By incorporating current, domain-specific knowledge, RAG minimizes the risk of hallucination, ensuring more reliable and context-aware solutions for precision agriculture. This innovative approach not only increases answer accuracy but also mitigates hallucinations, promising transformative impacts on agricultural practices.

Moreover, we wish to expand on our dataset and also prioritize meeting farmers and expert agriculturists interested in phytopathology researches. The cooperation among individuals and groups, Organizations and the government plays a pivotal role in advancing sustainable and effective agricultural systems, helping farmers worldwide, and improving agricultural practices. Furthermore, the extension of the model to include a broader range of disease classes in the two coffee species, Arabica and Robusta, is an avenue for future research.

Looking ahead, our work lays the groundwork for ongoing development. The focus on scalability, reliability, and user-friendly design positions this system as a practical tool for positive changes in agriculture. As we move forward, our goal is to expand the system's capabilities to address a wider range of diseases. This study, with its emphasis on simplicity and practicality, aims to contribute to the larger vision of sustainable and technologically enhanced food production.

\end{document}